\documentclass[12pt]{article}
\usepackage{amssymb,amsfonts}
\csname @addtoreset\endcsname{equation}{section}

\usepackage{graphicx}

\newcommand{\be}{\begin{equation}}
\newcommand{\ee}{\end{equation}}
\newcommand{\bea}{\begin{eqnarray}}
\newcommand{\eea}{\end{eqnarray}}
\newcommand{\bref}[1]{(\ref{#1})}

\newcommand{\G}{\Gamma} \newcommand{\D}{\delta} 
\newcommand{\ep}{\epsilon} \newcommand{\vep}{\varepsilon}
\newcommand{\T}{\theta}

\newcommand{\h}{\eta}

\newcommand{\Tb}{{\overline\theta}}

\def\6{\partial}
\def\7{\tilde}
\def\8{\widehat}

\def\pa{\partial}

\def\G11{\Gamma_{11} }
\def\NR{NR }
\def\lag{lagrangian }

\def\lag{lagrangian }

\begin{document}

\begin{titlepage}
\begin{flushright}
UB-ECM-PF-04/29\\KUL-TF-04/31
\end{flushright}
\vspace{1cm}
\begin{center}
\baselineskip=16pt {\LARGE    Rotating Solutions of Non-relativistic String
Theory}\\
\vfill
{\Large Joaquim Gomis $^{\dagger,1}$ and Filippo Passerini $^{\ast,2}$
  } \\
\vfill
{\small $^\dagger$  Departament ECM, Facultat de F{\'\i}sica, \\
Universitat de Barcelona and Institut de F{\'\i}sica d'Altes Energies, \\
Diagonal 647, E-08028 Barcelona, Spain \\
\vspace{0.5cm}
$^\ast$ Instituut voor Theoretische Fysica,\\ Katholieke Universiteit Leuven,\\
Celestijnenlaan 200D, B-3001 Leuven, Belgium}
\end{center}
\vspace{3cm}
\begin{center}
{\bf Abstract}
\end{center}
{\small We construct classical rotating solutions of Non-relativistic String
Theory.
The relation among the energy and angular momenta for these solutions
is of the type $E=J^2$. Some of the solutions  saturate a BPS bound for the
energy, they are ${1\over4}$ BPS supersymmetric configurations.


}
\vspace{2mm} \vfill
 \hrule width 7.3cm
 {\footnotesize \noindent
 $^1$  E-mail: gomis@ecm.ub.es\\
$^2$  E-mail: Filippo.Passerini@fys.kuleuven.ac.be}
\end{titlepage}


\section{Introduction}\label{sec.1}

Non-relativistic strings (\NR) in flat transverse space with a
non-trivial spectrum were introduced in
\cite{Gomis:2000bd}\cite{Danielsson:2000gi},
 in the static gauge they become  free theories
\cite{Garcia:2002fa}. The actions of \NR strings and branes
can be constructed as Wess-Zumino
terms \cite{Brugues:2004an}
 of the underlying Galilei groups using the method of
non-linear realizations for space-time groups \cite{Ivanov:1975zq}
\cite{Gauntlett:1990nk}.

\NR superstring theories in flat transverse space were recently
analyzed in \cite{Gomis:2004pw}. The action  was obtained by performing a
suitable non relativistic limit of Green-Schwarz type IIA string. These
\NR superstring theories have gauge diffeomorphism and kappa symmetry.

We hope the study of NR string theory could be usefull to find a new sector of
the theory where the AdS/CFT correspondence can be tested.

In this paper we consider \NR bosonic string in a curved transverse space
in terms of a Polyakov type action.
In the static gauge case, we
derive a BPS bound for the energy.
This theory  is conformal at quantum level
if the transverse metric is Ricci flat.

We study rotating solutions of the classical equations of motion
for the flat case and
for the singular conifold \cite{Candelas:1989js}. For the case of
the conifold the solutions we found
have three angular momenta. The relation among the energy and
angular momenta is of non-relativistic type $E\sim J^2$ where
$J$ represents the angular momenta.

We also analyze the supersymmetric properties of the bosonic
solutions we have found by studying the conditions imposed by
the kappa symmetry and supersymmetry of the \NR superstring
\cite{Gomis:2004pw}.

The organization of the paper is as follows. In section 2 we
introduce Polyakov form \NR string and derive a BPS bound for the
energy. In section 3 we find rotating solutions of \NR string in
flat space time. In section 4 we consider solutions in the case of
a singular conifold. In section 5 we study the supersymmetric
properties of our solutions. We also give some conclusions in
section 6.

\section{Non-relativistic strings in transverse curved  backgrounds
and BPS bound}
\label{sec.2}

We study a non-relativistic string in
a $d$-dimensional space-time, with coordinates $x^\mu$,
 $\mu=0,1$, along the string and transverse
coordinates $X^a$,  $a=2,\ldots,d-1$.
The $d$-dimensional metric is
 \begin{eqnarray}
  \label{metrica}
g_{MN}=\left(\begin{array}{cc}\eta_{\mu\nu}&0\\0&
G_{ab}(X^a)\end{array}\right).
\end{eqnarray}
 The $\eta_{\mu\nu}$ is the
$2$-dimensional Minkowski metric with signature $(-,+)$  and
$G_{ab}(X^a)$ is a metric on some $(d-2)$-dimensional Riemannian manifold
${\cal M}$.

The action of \NR string is given by
\be\label{azionephi}
S =  - {1\over2}T  \int d^{2}\sigma \sqrt{-\det g}\, g^{ij}
\partial_i X^a \partial_j X^b G_{ab}
\ee
where $g^{ij}$ is the inverse of the two dimensional metric
\be
g_{ij} = \partial_i x^\mu \partial_j x^\nu\eta_{\mu\nu},
\ee
the worldsheet coordinates are  $\tau$ and $\sigma$,
 $\sigma^i= (\tau, \sigma), i=1,2$.

The action is
invariant under 2d diffeomorphism of the world sheet.
In order to have a consistent non relativistic string theory at quantum level
we have to
consider the coordinate $x^1$ toroidally compactified
\cite{Gomis:2000bd}\cite{Danielsson:2000gi}, i.e.:
 \begin{equation}
 x^1\backsim x^1 + 2\pi R.
 \end{equation}

We now analyze the dynamics of non relativistic strings at
classical level by using
the Hamiltonian formalism.  Let us indicate by $p_\mu$ the
canonical momenta associated to the longitudinal coordinates $x^\mu$
 and $P_a$ the transverse momenta. As a consequence of the
gauge symmetry of the action (\ref{azionephi}), the canonical
variables satisfy the following two primary first class constraints:
\bea\label{constraints0}
V_0&=&p_\mu\;\vep^{\mu\rho}\eta_{\rho\nu}\;{x'}^\nu+
\frac12\left(\frac{P_aP_b}{T} G^{ab}(X)+{X'}^a{X'}^b  TG_{ab}(X)\right)
\sim 0,
\\\label{constraints1}
V_1&=&p_\mu {x'}^\mu+P_a{X'}^a\sim 0.
\eea

The  Dirac hamiltonian is:
\begin{equation}\label{hamdir}
 H_{D}=\int d\sigma
\left[
\lambda_{0} V_0+\lambda_{1} V_1\right]
\end{equation}
where $\lambda_{0},\lambda_{1}$
are arbitrary functions of the world sheet
coordinates $\sigma^i$.

If we write these arbitrary functions in terms of an auxiliary two
dimensional metric $\gamma_{ij}$,
$\lambda_{0}=\frac{\sqrt{-\gamma}}{\gamma_{11}},
\lambda_{1}=\frac{\gamma_{01}}{\gamma_{11}}$ we will find a gauge
invariant Polyakov's formulation of this non-relativistic string.
In fact if we consider the first order action \be {\tilde
S}_P=\int d^2\sigma \left[ p_\mu {\dot x}^\mu+P_a{\dot X}^a
\right]-\int d \tau H_D \ee and we eliminated the momenta $P_a$ we
get \be S_P=\int d^2\sigma \Bigg[-\frac{T}2 \sqrt{-\gamma}
\gamma^{ij}\partial_i X^a\partial_j X^b G_{ab}+ p_\mu(\dot
x^\mu-\frac{\sqrt{-\gamma}}{\gamma_{11}} \vep^{\mu\rho}\
\eta_{\rho\nu}\;{x'}^\nu-
\frac{\gamma_{01}}{\gamma_{11}}{x'}^\mu)\Bigg]. \label{polyakov}
\ee This action is  invariant under 2d diffeomorphism\footnote{We
acknowledge Kiyoshi Kamimura for useful discussions on this
point.} \bea
\delta x^\mu&=& \xi^k\partial_k x^\mu\\
\delta p_\mu&=& \partial_k(\xi^kp_\mu)+ p_\mu(-{\dot
\xi}^0+{\xi'}^0
\frac{\gamma_{01}}{\gamma_{11}})+p^\nu\vep_{\nu\mu}{\xi'}^0
 \frac{\sqrt{-\gamma}}{\gamma_{11}}\\
\delta X^a&=& \xi^k\partial_k X^a\\
\delta\gamma_{ij}&=& {\cal L}_\xi\gamma_{ij}
\eea
where ${\cal L}_\xi$ is the Lie derivative along $\xi$. The action is
also invariant under
gauge
Weyl transformations of the auxiliary metric $\gamma_{ij}$.

If we choose the conformal gauge $ \sqrt{-\gamma}
\gamma^{ij}=\eta^{ij}$, the action becomes
\be S_P=\int
d^2\sigma \left[-\frac{T}2 \eta^{ij}\partial_i
X^a\partial_j X^b G_{ab}+
p_\mu(\dot x^\mu-
\vep^{\mu\rho}\ \eta_{\rho\nu}\;{x'}^\nu)
 \right]. \label{polyakov1}
\ee

The equations of motion in the
conformal gauge are
\bea \label{eqmocur1}
0&=& \eta^{ij}\partial_i\partial_j X^a+\Gamma^a_{bc}
\eta^{ij}\partial_i X^b \partial_j X^c\\
0&=& \dot x^\mu-
\vep^{\mu\rho}\ \eta_{\rho\nu}\;{x'}^\nu\label{eqmocur1a}\\
0&=& \dot p_\mu- \vep^{\nu\rho}\ \eta_{\rho\mu}\;{p'}_\nu \eea
where $\Gamma^a_{bc}$ are the Christoffel symbols of the metric
$G_{ab}$. From \bref{eqmocur1a} we deduce that the longitudinal
coordinates verify the relations $\dot x x'=0$ and ${\dot
x}^2+{x'}^2=0$.

The boundary conditions are the same as the ones considered for the flat space
case \cite{Garcia:2002fa}, i.e. for  a closed string
\begin{equation}
\label{chiuse} x^\mu(\tau,\sigma+2\pi)=x^\mu(\tau,\sigma)+
2\pi nR\delta^\alpha_1  \qquad
X^a(\tau,\sigma+2\pi)=X^a(\tau,\sigma)
\end{equation}
where $n\in \mathbb{Z}$ is the winding number of the string.\\
For an open string
\begin{eqnarray} \label{aperte}\nonumber
x'^0|_{\sigma=\pi}=x'^0|_{\sigma=0}=0\qquad
x^1(\sigma=0)=0 \qquad x^1(\sigma=\pi)=2\pi nR \\
X'^a_N|_{\sigma=\pi}=X'^a_N|_{\sigma=0}=0 \qquad
\dot{X}^b_D|_{\sigma=\pi}=\dot{X}^b_D|_{\sigma=0}=0,
\end{eqnarray}
also in this case  $n\in \mathbb{Z}$ .

If we introduce the coordinates
$\gamma=X^0+X^1,\bar\gamma=X^0-X^1, \beta=p_0+p_1,
\bar\beta=p_0-p_1$ the first order part of the action
\bref{polyakov1} is the conformal (2,2) $\beta\gamma$ system
introduced in \cite{Gomis:2000bd}. Therefore  our NR string will
be conformal invariant at quantum level if the  metric $G_{ab}$ is
Ricci flat \cite{Callan:1985ia} and the spacetime dimension is 26
\cite{Gomis:2000bd} like for the ordinary bosonic relativistic
string.

To solve the classical equations of motion is convenient to work in the
static gauge. We fix this gauge by imposing two gauge fixing
constraints:
\begin{eqnarray}\label{statclos}
\Phi_0\equiv x^0-K\tau=0 \qquad \Phi_1\equiv x^1-K\sigma=0
\end{eqnarray}
where $K$ is a constant. These constraints make the constraints
\bref{constraints0} \bref{constraints1} second class, they are stationary
if $\lambda_{0}=1, \lambda_{1}=0$. Therefore the static
gauge is included in the conformal gauge. In the static gauge  the
transverse degrees of freedom, $(X^a, P_a)$, are the
independent  degrees of freedom of the \NR string.
The dynamics is given by \bref{eqmocur1}.

 The momenta ${\cal P}_1$ along the string,
in the static gauge, using the constraints \bref{constraints1}
is given by
\be\label{momenta1}
{\cal P}_1=-\frac{1}{K}\int d\sigma\left(P_a{ X'}^a \right)
\ee
and the energy is
\begin{equation}
\label{energia} E=\frac{1}{K}\int
d\sigma\left(\frac{(P)^2}{2T}+ \frac{T(X')^2}{2}\right)
\end{equation}
where we have used the constraint \bref{constraints0}. Note that
the energy can be written as \be\label{relb} E=\int d\sigma\frac
{(P^a\pm TX'^a)^2}{2TK}\pm {\cal P}_1 =\int d\sigma\frac {T(\dot
X^a\pm X'^a)^2}{2K}\pm {\cal P}_1.\ee

Therefore ${\cal P}_1$ is the BPS bound
\be
\label{bound} E\ge \pm {\cal P}_1\ge
0,
\ee
the bound is saturated by the BPS configurations \be\label{bpseq}
\dot X^a\pm X'^a=0. \ee
The configurations that satisfy the BPS equation \bref{bpseq}
verify also the second order equations of
motion \bref{eqmocur1}.

\section{Classical Solutions for NR strings in flat space time}
We now analyze non relativistic strings in flat transverse space,
i.e. we consider $G_{ab}(X^a)=\delta_{ab}$. The equations of
motion (\ref{eqmocur1}) simplify to:
\begin{equation}
 \label{onda} \ddot{X}^a=X''^a.
\end{equation}
Here we are interested in solutions of (\ref{onda}) describing
rotating strings, i.e.  solutions with angular momentum different
from zero. The component of the angular momentum perpendicular to
the $X^a-X^b$ plane is:
\begin{equation} \label{mom}
J^{ab}=\int d\sigma\left(X^aP^b-X^bP^a\right).
\end{equation}

\subsection{Closed Strings}
We now consider solutions of the equations (\ref{onda}) satisfying
boundary conditions for closed strings (\ref{chiuse}). To satisfy
the boundary conditions for  the longitudinal coordinates we have
to fix $K=nR$ in the static gauge constraints (\ref{statclos}).

\begin{figure}
  \centerline{\includegraphics[width=10cm]{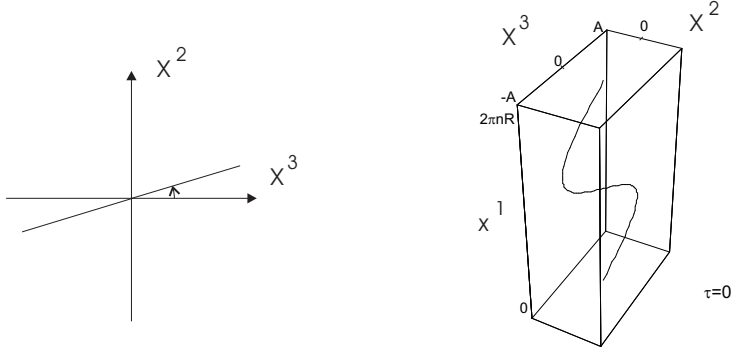}}
\caption{String described by solution (\ref{soluzione1}) when
$\omega=1$}\label{f12}
\end{figure}

The  simplest solution describing a rotating closed string is:
\begin{eqnarray}\nonumber
x^0=nR\tau \qquad
X^2=A\sin(\omega\sigma)\sin\left(\omega\tau\right)\\\label{soluzione1}
x^1=nR\sigma\qquad
X^3=A\sin(\omega\sigma)\cos\left(\omega\tau\right)
\end{eqnarray}
where $\omega\in\mathbb{Z}$ and $A$ is a dimensional constant,
$[A]=L$; these conventions are  valid also for the following
solutions. The (\ref{soluzione1}) describes a circular string with
angular momentum perpendicular to the $X^2-X^3$ plane (figure
\ref{f12}).
The velocity modulus of the points of the string is:
\begin{eqnarray}
 v^2=\left(\frac{\partial X^2}{\partial
x^0}\right)^2+\left(\frac{\partial X^3}{\partial
x^0}\right)^2=\left(\frac{\omega
A}{nR}\right)^2\left(\sin(\omega\sigma)\right)^2\qquad
v_{max}=\frac{A\omega}{nR} \qquad\label{speed1}.
\end{eqnarray}

Like for the relativistic case the turning points are the faster
points, but in this case the maximum velocity is not limited by
the velocity of light, which in natural units is $v=1$.
This is a typical feature of non
relativistic theory. By using (\ref{energia}) and (\ref{mom}) we
find:
\begin{equation}
\label{rels} E=\frac{J^2}{TA^2\pi nR}.
\end{equation}
Interpreting  $\frac{1}{2}TA^2\pi nR$ like the inertial
momentum of the string respect to the rotation axis,  the
(\ref{rels}) is the dispersion relation for a classical rotating
system.

The momentum along the string ${\cal P}_1$ vanishes for this configuration.
Since the energy of this configuration \bref{rels} is different from zero
this configuration does not saturate the bound of the energy \bref{bound}
and therefore we expect it will be non-supersymmetric if we embed this
\NR string in a supersymmetric theory.

A generalization of this solution with more than two spikes in the
transverse space  is given by
\begin{eqnarray}
\nonumber x^0=nR\tau\qquad
X^2=B\left(\cos((z-1)(\tau+\sigma))+(z-1)\cos(\tau-\sigma)\right)
\\\label{spike} x^1=nR\sigma\qquad
X^3=B\left(\sin((z-1)(\tau+\sigma))+(z-1)\sin(\tau-\sigma)\right),
\end{eqnarray}
where z is an integer and corresponds to the number of spikes of the strings.
When $z=2$, we recover the solution \bref{soluzione1} with
$A=2B$ and $\omega=1$.
These solutions are
the NR version of the solutions studied in  \cite{Kruczenski:2004wg}.

The spikes are the fastest point of the strings, the modulus of their velocity
is
\bea
v=\frac{2B(z-1)}{nR}\label{spikevel}\eea
i.e. a non-limited quantity. Using (\ref{energia}) and (\ref{mom}) we
find:
\begin{equation}
 E=\frac{J^2}{TB^2\pi nRz^2}.
\label{spikeene}\end{equation}
The inertial momentum $\frac{1}{2}TB^2\pi nRz^2$ is proportional to the square
of the number of spikes. This is in agreement with the fact that increasing the
number of spikes, the string is on average more distant from the rotational
axis.
For these solutions also ${\cal P}_1=0$ and the BPS bound is not saturated.

\vspace{5mm}

We now analyze a solution describing a string that is moving in a
$s$-dimensional sphere  embedded in the transverse space. The
transverse part is described by a $(s+1)$-dimensional vector of
solutions \cite{Hoppe:1987vv}:
\begin{eqnarray}
\label{hoppe}
\overrightarrow{X}(\tau,\sigma)=D(\tau)\overrightarrow{m}(\sigma)
\end{eqnarray}
where:
\begin{eqnarray}
\overrightarrow{m}(\sigma)=(A_1\sin(\omega_1\sigma),A_1\cos(\omega_1\sigma),
A_2\sin(\omega_2\sigma),A_2\cos(\omega_2\sigma),\ldots)
\\\nonumber\\D(\tau)=\left(\begin{array}{ccccc}\sin\left(\omega_1\tau\right)&
\cos\left(\omega_1\tau\right)&0&0&\ldots\\\\
\cos\left(\omega_1\tau\right)&-\sin\left(\omega_1\tau\right)&0&0&\ldots\\\\
0&0&\sin\left(\omega_2\tau\right)&\cos\left(\omega_2\tau\right)&\ldots\\\\
0&0&\cos\left(\omega_2\tau\right)&-\sin\left(\omega_2\tau\right)&\ldots\\\\
\vdots & \vdots &\vdots & \vdots &\ddots
\end{array}\right).
\end{eqnarray}
This solution verify the BPS equation \bref{bpseq},
$\dot{\vec X}+ \vec X'=0$,
and therefore
satisfy the equations of motion
(\ref{onda}) with the boundary
  conditions (\ref{chiuse}). This solution describe
a string on a $s$-dimensional sphere since
\begin{eqnarray}
\label{sferas}\overrightarrow{X}(\tau,\sigma)^2=\textrm{constant}.
\end{eqnarray}

\begin{figure}
\center{\includegraphics[width=10cm]{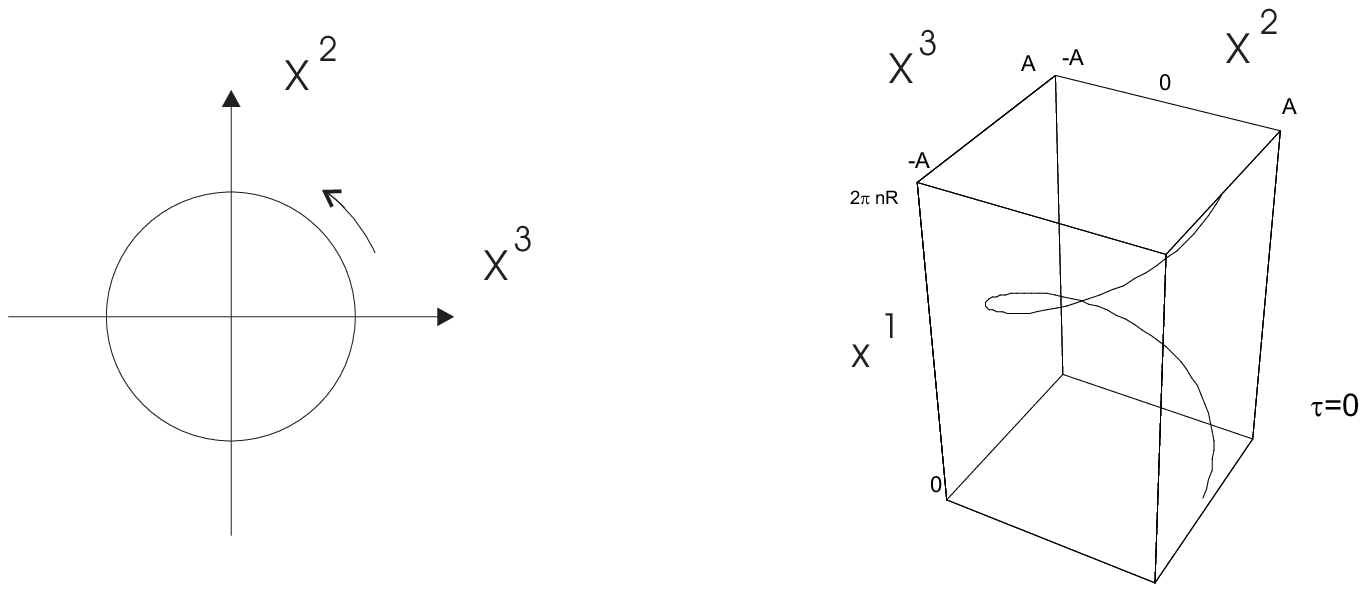}}
\caption{String described by solution (\ref{semhoppe}) when
$\omega=1$}\label{f34}
\end{figure}

The  simplest solution of this kind, considering also the
longitudinal part, is:
\begin{eqnarray}\label{susysol}
\nonumber x^0=nR\tau\qquad
X^2=A\left(\sin(\omega\sigma)\sin\left(\omega\tau\right)+\cos(\omega\sigma)\cos\left(\omega\tau\right)\right)
\\\label{semhoppe} x^1=nR\sigma\qquad
X^3=A\left(-\cos(\omega\sigma)\sin\left(\omega\tau\right)+\sin(\omega\sigma)\cos\left(\omega\tau\right)\right).
\end{eqnarray}
The solution (\ref{semhoppe}) describes a  circular string
with angular momentum perpendicular to the plane $X^2$-$X^3$,
(figure \ref{f34}).
The velocity modulus is the same for all the points of the string:
\begin{equation}
v=\frac{A\omega}{nR},
\label{speed2}
\end{equation}
also in this case there is not a limit for the velocity. From
(\ref{energia}) and (\ref{mom}) we have:
\begin{equation} \label{rel}
E=\frac{J^2}{2TA^2\pi nR}.
\end{equation}
For this solution the inertial momentum is $TA^2\pi nR$ and is
bigger than that one of the  solution \bref{soluzione1}. This is
in agreement with the classical intuition because now the string
is on an average more distant from the rotation axis.

For this solution the momenta along the string is given by \be
{\cal P}_1=\frac{J^2}{2TA^2\pi nR} \ee and coincides with the
energy of the solution because this configuration verifies the BPS
equation and therefore it saturates the BPS bound. We will see
that this configuration is ${1\over 4}$ BPS supersymmetric when we
embed the bosonic \NR string in the \NR superstring
\cite{Gomis:2004pw}.

Relativistic  closed string   on spheres with all points moving at
the same velocity in modulus, was considered in
\cite{Hoppe:1987vv} .
They have at least
two component of the angular momentum different from zero. However
 they are  different of our \NR solutions among other things
because they are not supersymmetric.

 \subsection{Open strings}
\begin{figure}
  \centerline{\includegraphics[width=10cm]{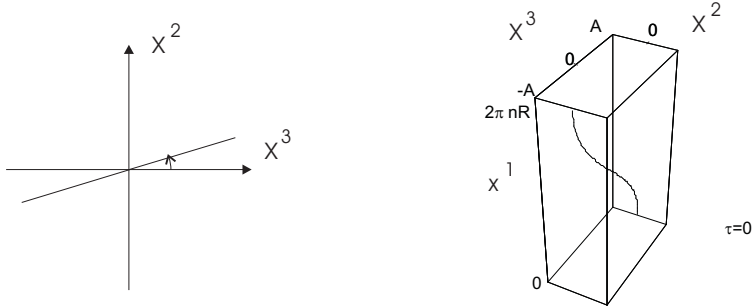}}
\caption{String described by solution  (\ref{san}) when
$\omega=1$}\label{an}
\end{figure}

We now turn to  open strings, i.e.  solutions of the equations
(\ref{onda}) satisfying boundary conditions (\ref{aperte}). To
satisfy the boundary conditions for the longitudinal coordinates
we have to fix $K=2nR$ in the static gauge constraints
(\ref{statclos}). We consider a solution with all transverse
coordinates satisfying Neumann boundary conditions:
\begin{eqnarray}
\nonumber x^0=2nR\tau \qquad
X^2=A\cos(\omega\sigma)\sin\left(\omega\tau\right)\\\label{san}
x^1=2nR\sigma\qquad
X^3=A\cos(\omega\sigma)\cos\left(\omega\tau\right).
\end{eqnarray}
This is an open string with angular momentum perpendicular to the
$X^2-X^3$ plane (figure \ref{an}).
The string extrema are the faster points and also in this case
 the velocity modulus $v_{max}=\frac{A\omega}{2nR}$ in not limited.
 The relation between energy and angular momentum is
 \begin{equation}
\label{rela} E=\frac{J^2}{TA^2\pi nR},
 \end{equation}
 i.e. the same relation that we have obtained for the first kind of closed string.
 For relativistic strings there are different  energy-angular momentum relations for
  open and closed strings.

\section{Classical Solutions for NR strings in curved space time}
We now analyze the dynamics of strings propagating in a curved
transverse space. Since the transverse metric should be Ricci flat
in order to have a consistent \NR string at quantum level, here we
consider  the case of a singular conifold with metric
$ds^2=dr^2+r^2 ds^2_{T^{1,1}}$. The $T^{1,1}$ is the conifold
base, i.e.  the 5-dimensional space
$T^{1,1}$\cite{Candelas:1989js},
\begin{eqnarray} \label{(T)}\nonumber
ds^2_{(T^{1,1})}=\qquad\qquad\qquad\qquad\qquad\qquad\\=\left(\frac{1}{9}(d\psi+\cos\theta_1
d\phi_1+\cos\theta_2
d\phi_2)^2+\frac{1}{6}(d\theta_1^2+\sin^2\theta_1
d\phi_1^2+d\theta_1^2+ \sin^2\theta_2 d\phi_2^2)\right)\qquad
\end{eqnarray}
 where  the coordinates
$X^a$ are renamed $r, \psi$, $\phi_1$,
$\phi_2$, $\theta_1$, $\theta_2$. This metric does not depend on $\psi$,
$\phi_1$, $\phi_2$  and the action is invariant under
translations of these angles. The conserved quantities
associated to these symmetries  are the angular momenta:
\begin{eqnarray}
\label{momcurv}\nonumber J_i=\int^{2\pi}_{0}d\sigma
P_{\hat a}=T\int^{2\pi}_{0}d\sigma\dot{X^{\hat b}}
G_{\hat a \hat b}^{(T^{1,1})}\qquad
\hat a, \hat b=1,2,3 \\\qquad(X^1=\psi\qquad X^2=\phi_1\qquad
X^3=\phi_2)\qquad
\end{eqnarray}

We now consider solutions of the equations of motion
(\ref{eqmocur1}) describing closed strings, i.e. solutions
satisfying the boundary conditions (\ref{chiuse}).
\subsection{Point like strings}
We write three different solutions describing strings that in the
transverse space are point particle periodically moving.
Considering also the longitudinal part, these are circular
extended strings. For these solutions the energy is obtained by
(\ref{energia}) and the angular momentum by (\ref{momcurv}). All
of them have $r=C$, where $C$ is a constant:
\begin{itemize}
\item One $\tau$-dependent transverse coordinate:
\begin{eqnarray}
 \label{punto1}
  x^0=nR\tau\qquad x^1=nR\sigma\qquad
\psi=\nu nR\tau \qquad\phi_1=\phi_2= \theta_1=\theta_2=0
\qquad\end{eqnarray}
\begin{eqnarray}
\label{mompunt} J_\psi=J_{\phi_1}=J_{\phi_2}=TnR2\pi\nu
\frac{C^2}{9} \qquad\qquad E=nRT\pi\nu^2\frac{C^2}{9}
\end{eqnarray}

\item Two $\tau$-dependent transverse coordinates:
\begin{eqnarray} \label{punto2}
  x^0=nR\tau\qquad x^1=nR\sigma\qquad
\psi=\phi_1=\nu nR\tau \qquad \theta_1=\theta_2=\phi_2=0\qquad
\end{eqnarray}
\begin{eqnarray}
\label{mompunt} J_\psi=J_{\phi_1}=J_{\phi_2}=TnR2\pi\nu
\frac{2}{9}C^2 \qquad\qquad E=nRT\pi\nu^2\frac{4}{9}C^2
\end{eqnarray}

\item Three $\tau$-dependent transverse coordinates:
\begin{eqnarray}
 \label{punto3}
  x^0=nR\tau\qquad x^1=nR\sigma\qquad
\psi=\phi_1=\phi_2=\nu nR\tau \qquad \theta_1=\theta_2=0\qquad
\end{eqnarray}
\begin{eqnarray}
\label{mompunt} J_\psi=J_{\phi_1}=J_{\phi_2}=TnR2\pi\nu
\frac{C^2}{3}\qquad\qquad  E=nRT\pi\nu^2C^2
\end{eqnarray}
\end{itemize}

For the three point like solutions, the   relation between energy
and angular momentum is the same, i.e.:
\begin{equation} \label{enpunt}
E=3\frac{J_\psi^2}{4\pi nRTC^2}+3\frac{J_{\phi_1}^2}{4\pi
nRTC^2}+3\frac{J_{\phi_2}^2}{4\pi nRTC^2}.
\end{equation}

For all these point like solutions the momenta ${\cal P}_1=0$.

\subsection{Extended string}
We now consider a string  that is extended also  in the transverse space and  like
\bref{semhoppe} is moving along its own extension:
\begin{eqnarray}\label{3.10like}
x^0=nR\tau\qquad x^1=nR\sigma\qquad\phi_1=\nu
nR\tau-\omega\sigma\qquad\psi=\phi_2=\theta_1=\theta_2=0\qquad
\end{eqnarray}
where $\omega\in\mathbb{N}$. This is a closed
 string because
 $\phi_1\in(0,2\pi)$.

>From (\ref{momcurv}) we obtain:
\begin{eqnarray}
 J_\psi=J_{\phi_1}=J_{\phi_2}=TnR2\pi\nu \frac{C^2}{9}.
\end{eqnarray}

Using   (\ref{energia}) we have:
\begin{eqnarray}\label{en3.10like}
E=nRT\pi\nu^2\frac{C^2}{9}+\pi\frac{C^2}{9}\frac{T}{nR}\omega^2
\end{eqnarray}
and  finally we find
\begin{equation} \label{enext1}
E=3\frac{J_\psi^2}{4\pi nRTC^2}+3\frac{J_{\phi_1}^2}{4\pi
nRTC^2}+3\frac{J_{\phi_2}^2}{4\pi
nRTC^2}+\pi\frac{C^2}{9}\frac{T}{nR}\omega^2.\qquad\end{equation}

For this solution \be\label{P3.10like} {\cal
P}_1=T\nu\omega2\pi\frac{C^2}{9}.\ee

The BPS equation  \bref{bpseq} is satisfied when $\nu=\frac{\omega}{nR}$; in fact,  in this case from
\bref{en3.10like}  and  \bref{P3.10like} we have $E={\cal P}_1$.

\section{Supersymmetry properties of the solutions}
In \cite{Gomis:2004pw} it has been considered a
supersymmetric extension of the bosonic
string \bref{azionephi} in flat space time.
The \NR superstring has diffeomorphism and kappa invariance in 10 dimensions.
The action in the static gauge and with half of the fermions
set to zero has the following form
\be\label{susyaction}
S= -T\;\int d^2\sigma \left[ \frac{1}2\;\h^{ij} \pa_i
X \cdot \pa_j X \;+\;2
i\; \Tb_+\Gamma^i\pa_i\T_+\right]
\label{Lag22gf2}
\ee
where $\theta_+$ is a $+1$ eigenspinor of $\Gamma_*=
\Gamma_0\Gamma_1\Gamma_{11}$ with 16 components and
$\bar\theta_+$ is the conjugate
spinor.
Note that this \lag is quadratic in the bosonic and
fermionic variables.

The action \bref{Lag22gf2} is invariant under the supersymmetry
transformations
\be
\D\T_+= \ep_+\;+\;\frac{1}{2}\Gamma_\mu\partial_i x^\mu\pa^i X^a
\Gamma_a \ep_-,\quad \quad
\D X^a\;=\;2i\Tb_+\Gamma^a\ep_-
\label{ressusyst}
\ee
where  $\ep_+$ and $\ep_-$ are 16 components constant spinors.

The bosonic supersymmetric configurations of the action \bref{susyaction}
should verify
\be
\ep_+\;=\;-\frac{1}{2} \Gamma_\mu\partial_i x^\mu\pa^i x^a\Gamma_a \ep_-,
\label{susycond}
\ee
with $ \ep_\pm$ constants spinors. When
 this relation is verified for some non-vanishing
components of $\ep_\pm$ the configuration preserves some of the 32
supersymmetries of the action \bref{Lag22gf2}.

The  vacuum configuration of the string
\be\label{vacuum}
x^0=K\tau,  \quad \quad x^1=K\sigma, \quad \quad \;X^a= {\rm constant}
\ee
is supersymmetric if $\ep_+\;=\;0$ which implies
\be
\Gamma^0\Gamma^1\G11\ep=\ep.
\label{gammacond2}
\ee
Therefore the string is a $\frac12$ BPS configuration.
Note that the vacuum solution \bref{vacuum} is a solution of both
the relativistic and \NR string.

There are other possible bosonic supersymmetric configurations
with non-constant transverse coordinates. If we write
$X^a=X^a(\tau,\sigma)$, we can have a solution of \bref{susycond} if
\be\label{susyconn}
\partial_\tau X^b=\pm\partial_\sigma X^b,\quad\quad {\rm and}\quad\quad
\Gamma_0\ep_-=\pm\Gamma_1\ep_-.
\label{quarters2}\ee
Together with the condition $\ep_+=0$ the susy parameter $\ep$
should satisfy \be \mp\Gamma^0\Gamma^1\ep=\ep. \label{quarters}
\ee

This configuration is a ${1\over4}$ BPS configuration. It
represents a wave propagating, with the velocity of light, along a
string with arbitrary profile in the transverse directions.

The supersymmetric (BPS) condition \bref{susyconn} was obtained previously as
 a condition to saturate the energy bound \bref{bound}.

Now we can study the symmetry properties of the solutions we have
found. The solutions   \bref{soluzione1} and \bref{spike} are not
supersymmetric because the condition \bref{susycond} can not be
verified. Instead the  solution   \bref{susysol} verifies the
supersymmetry condition if the spinors verify $\ep_+\;=\;0$ and
$\Gamma_0\ep_-=\pm\Gamma_1\ep_-$. It is therefore a ${1\over4}$
BPS configuration. For the solutions in flat space time that we
have considered, we observe that if the BPS energy bound \bref{bound} is verified the
solution is supersymmetric.

In the case of a transverse curved manifold there is not yet a
general discussion about the supersymmetry properties of a
\NR superstring in these backgrounds. Therefore for
the solutions we have found in transverse curved background
we make the ansatz that the solutions
that saturate the energy  bound \bref{bound} will be supersymmetric.

\section{Conclusions}
In this   paper we have  considered  NR  string theory in curved
transverse space. Our starting point has been the Nambu-Goto  action for
NR strings
in a curved  transverse space \bref{azionephi}.
We have then  rewritten the action in  a
Polyakov like form \bref{polyakov}
and we have derived a BPS bound for the energy in the static gauge.

We have seen that the theory is
conformal at quantum level if the transverse space is Ricci flat.
For this reason we have considered the case of a singular conifold
\cite{Candelas:1989js}.

We  also have found  classical rotating solutions of  NR
strings. The solutions have
typical features of a non relativistic theory. In
particular, the velocity of the strings has not any  limit
\bref{speed1},\bref{spikevel},\bref{speed2} and the energy is proportional to
squared angular momentum
\bref{rels},\bref{spikeene},\bref{rel},\bref{rela},\bref{enpunt},\bref{enext1}.

For the flat space case, the BPS equation is  obtained as condition to saturate
the energy bound \bref{bpseq} or as a supersymmetry condition
\bref{susyconn}.

Some of the solutions are ${1\over4}$ BPS supersymmetric and represent
a wave with the velocity of light propagating along the string.

These \NR string have a energy spectrum of non-relativistic theories
but keep  relativistic properties along the longitudinal directions.

\section{Acknowledgements}
We are grateful to Roberto Emparan, Jaume Gomis, Kiyoshi Kamimura,
Tomas Ort\'{\i}n, Josep Maria Pons,
 Paul Townsend and Toine Van Proeyen for useful discussions and
 comments.
This work is supported in part by the European Community's Human
Potential Programme under contract MRTN-CT-2004-005104 `Constituents,
fundamental forces and symmetries of the universe'.
The work of F.P. is supported in part  by the Federal Office for
Scientific, Technical and Cultural Affairs through the "Interuniversity
Attraction Poles Programme -- Belgian Science Policy" P5/27.

\end{document}